Coexistence of competing orders with two energy gaps in real and momentum space in high-$T_c$ superconductor $Bi_2Sr_{2-x}La_xCuO_{6+\delta}$

J.-H. Ma,[1,*] Z.-H. Pan,[1,*] F. C. Niestemski,[1] M. Neupane,[1] Y.-M. Xu,[1] P. Richard,[2,1] K. Nakayama,[3] T. Sato,[3] T. Takahashi,[2,3] H.-Q. Luo,[4] L. Fang,[4] H.-H. Wen,[4] Ziqiang Wang,[1] H. Ding[1,4,†] and V. Madhavan[1,†]

[1]Department of Physics, Boston College, Chestnut Hill, MA 02467, USA

[2]WPI Research Center, Advanced Institute for Materials Research, Tohoku University, Sendai 980-8577, Japan

[3]Department of Physics, Tohoku University, Sendai 980-8578, Japan

[4]Institute of Physics and National Laboratory for Condensed Matter Physics, P. O. Box 603 Beijing, 100080, P. R. China

The superconducting phase of the high-$T_c$ cuprates has been thought to be described by a single d-wave pairing order parameter. Recently, there has been growing evidence suggesting that another form of order, possibly inherited from the pseudogap phase above $T_c$, may coexist with superconductivity in the underdoped regime. Through a combined study of scanning tunneling microscopy and angle-resolved photoemission spectroscopy, we report the observation of two distinct gaps (a small-gap and a large-gap) that coexist both in real space and in the anti-nodal region of momentum space in the superconducting phase of $Bi_2Sr_{2-x}La_xCuO_{6+\delta}$. We show that the small-gap is associated with superconductivity. The large-gap persists to temperatures above the transition temperature $T_c$ and is found to be linked to short-range charge ordering. Remarkably, we find a strong, short-ranged correlation between the local small- and large- gap magnitudes suggesting that the superconductivity and charge ordering are affected by similar physical processes.

\* These authors contributed equally to this work

† To whom correspondence should be addressed: dingh@bc.edu, madhavan@bc.edu

One of the central problems in high-$T_c$ superconductivity is understanding the origin of the pseudogap phase which manifests itself as a leading-edge spectral gap surviving above the transition temperature, $T_c$[1,2]. The crux of the intense debate on the pseudogap revolves around the issue of whether it is a precursor pairing state without superconducting coherence or a competing phase with a hidden order parameter. Many earlier experimental results on the pseudogap demonstrated characteristics of a precursor pairing gap, including similar gap amplitudes and d-wave-like momentum dependence above and below $T_c$, and smooth temperature evolution through $T_c$[3]. This has led to the belief that the superconducting phase is characterized by a single d-wave pairing order parameter which finds support in the observation of a single d-wave gap function in cases where the Bogoliubov quasi-particle peak survives at the antinode[4]. Recently, there has been increasing evidence for the existence of two distinct gaps associated with different order parameters in the superconducting phase below $T_c$, such as deviations from a single d-wave gap function[5,6], opposite doping dependence for the two gaps[7,8], and different temperature dependences of the two gaps[4,5,9], all of which suggest a competing order explanation for the pseudogap phase.

In this paper we report a detailed study of lanthanum substituted $Bi_2Sr_{2-x}La_xCuO_{6+\delta}$ (La-Bi2201), using the complementary experimental probes of scanning tunneling microscopy (STM) and angle-resolved photoemission spectroscopy (ARPES). STM data were obtained at 5K on samples that were cleaved in UHV and immediately inserted into the cold STM head described elsewhere[10]. ARPES experiments were performed at the Synchrotron Radiation Center in Wisconsin, and in the ARPES laboratory of Tohoku University using a microwave-driven Xenon source (hν = 8.437 eV)[11]. High-resolution ($E_{Re}$ < 4 meV, $k_{Re}$ < 0.005 Å$^{-1}$) was achieved by using this low photon energy.

We focus on two dopings: nearly optimally doped $Bi_2Sr_{1.6}La_{0.4}CuO_{6+\delta}$ (0.4La) with a $T_c$ of 32 K and over-doped $Bi_2Sr_{1.9}La_{0.1}CuO_{6+\delta}$ (0.1La) with a $T_c$ of 16K. Figures 1, A and E show STM *dI/dV* spectra on 0.4La and 0.1La samples. In the nearly optimally doped 0.4La samples (the 0.1La data will be discussed later), we observe two distinct but spatially coexisting gaps at low temperatures, which will from now on be referred to as small gap ($\Delta_s$) and large gap ($\Delta_L$). As seen in the linecut (Fig. 1D), both $\Delta_s$ and $\Delta_L$ vary with spatial location. This is illustrated statistically by the histograms of the two gaps obtained from dI/dV maps (Fig. 1C) that show a clear bimodal distribution of gap values. The average values of $\Delta_s$ and $\Delta_L$ over maps taken in different regions are 11.4 ± 4 meV and 33 ± 10 meV respectively.

For a comparison of real space and momentum space (k-space) data, high resolution (<4 meV) APRES data were obtained on both 0.4La and 0.1La samples. These complementary data sets on identical samples provide important new information that cannot be obtained from one technique alone. Low temperature (5K) ARPES spectra on the 0.4La samples show an angle dependent gap (Fig. 2, A and D) continuing smoothly into the antinodal region ($\theta = 0°$), albeit with suppressed quasiparticle coherence peaks. As seen in figure 2D, a d-wave function fits most of the data points from the node ($\theta = 45°$) to the coherent arc tip (defined as the point in k-space beyond which the coherence peaks are largely suppressed) indicating that they represent gaps from one order parameter. We find that the average STM small gap (11.4 meV) is comparable to the low temperature ARPES gap (Fig. 1, B and C) at the arc tip (~11meV). Coupling this with our finding that the ARPES gap near the nodal region disappears above $T_c$ (Fig. S1) allows us to associate the small STM gap with superconductivity. This immediately leads us to the question, is there a counterpart of the STM large gap in ARPES?

Earlier ARPES experiments on similar samples (Pb-La-Bi2201) found a large antinodal gap (pseudogap) of about 35meV both above and below $T_c$[6], similar in magnitude to our STM large gap. Thus, one could invoke the idea that there are two gaps which coexist spatially but occupy different regions of momentum space (solid lines in Fig. 2F). Within this picture, the large gap would be responsible for truncating the antinodal part of the Fermi surface and producing the Fermi arc in the normal state pseudogap phase, while pairing below $T_c$ is restricted to the Fermi arc. Remarkably, we find that this picture may be incorrect. Our low photon energy, high resolution ARPES enhances the features of the coherent excitations such that even data near the antinodal region (Fig. 2D) reveal smaller gaps (maximum gap ~14 meV) which follow the d-wave fit reasonably well. This indicates that the superconducting gap persists to the antinodes, which is also consistent with recent ARPES experiments on optimally doped (34K) La-Bi2201[12]. The observed "pairing beyond the Fermi arc" can be explained by the fact that the large pseudogap near the antinodes[1,6] is a soft gap with suppressed but finite in-gap density of states that could still facilitate pairing below $T_c$.

The above discussion suggests that the various experimental data can be reconciled within a scenario where the small and large gaps both coexist in the antinodal region[13]. The large gap near the antinodes is not clearly visible in our low temperatures ARPES spectra since the spectral weight associated with the superconducting coherence masks the signal from the large gap. Accordingly, we expect the larger gap to come to the forefront after the coherence peaks disappear above $T_c$. Indeed, this is precisely what we observe. ARPES antinodal spectra at higher temperatures show a distinctly larger pseudogap which eventually disappears at approximately 180K as shown in Fig. 2B. Furthermore, a division of the 5K spectrum by the 40K spectrum (Fig. 2E) also shows exactly the effect we expect, i.e., the coherence peaks that mask the larger gap are now clearly visible. Interestingly, the recent ARPES data on optimally doped La-Bi2201 show a squeezed "peak-dip-hump" feature in the antinodal region[12]. Since the hump has the correct energy scale to be the manifestation of the antinodal larger gap, this peak-dip-hump feature could also be attributed to two gaps. Finally, we would like to emphasize that based on our data the two gaps remain distinct and do not merge into one `quadrature' gap, ($\Delta = \sqrt{\Delta_s^2 + \Delta_l^2}$). Instead, our data are consistent with the new picture where below $T_c$, the superconducting gap persists beyond the arc tip all the way to the antinode where it coexists simultaneously with the larger gap (dotted lines in Fig. 2F). This new picture is also consistent with the recent theoretical work where the superconducting state coexists with glassy charge density wave order[14].

STM and ARPES on the overdoped 0.1La samples take us a step further in our understanding. ARPES data at 5K (Fig. 1F and fig. 2C) reveal that the superconducting gap and coherence persist deep into the antinodal region. Temperature dependence of the ARPES spectra show that the gap along much of the Fermi surface (75° > θ > 15°) vanishes above $T_c$. But surprisingly, the antinodal gap survives above $T_c$ with a similar gap magnitude but without coherence peaks (Fig. 2C). This gap vanishes between 80K and 100K, indicating a persisting pseudogap in the overdoped region consistent with the conclusion of a prior STM study on overdoped Pb-Bi2201 samples[15]. Correspondingly, our STM data reveal two spatially coexisting gaps (Fig. 1, G and H) confirming the presence of the pseudogap below $T_c$ and providing further evidence for the coexistence of the superconducting gap and pseudogap of closer values near the antinodal region in the overdoped samples. Comparing the STM and ARPES gaps below $T_c$ for 0.1La, we find that the average small gap (7 ± 2 meV) is comparable to the antinodal ARPES gap (~ 8 meV). Thus for both dopings the average STM small gap can be identified with the superconducting order parameter and scales as expected with $T_c$.

Having identified $\Delta_s$ with superconductivity, what can we say about the origin of the large gap, $\Delta_L$? There are many conjectures for the pseudogap state in hole-doped cuprates ranging from static order to fluctuating order of density waves in the particle-particle or the particle-hole channels. Charge ordering in the form of short-range ordered checkerboard patterns has indeed been previously observed by STM below $T_c$ in underdoped $Bi_2Sr_2CaCu_2O_{8\pm\delta}$ (Bi2212) [16] and Ca-oxychloride [17] as well as above $T_c$ in marginally underdoped Bi2212[18] and most recently in Pb-La-Bi2201[19]. To explore the possibility of a co-existing phase in the charge channel, we obtain Fourier transforms of *dI/dV* maps at various energies. On the 0.4La samples, a glassy form of charge order revealed as shown in Fig. 3A. The corresponding wave vector (q-vector) observed in the Fourier transform is $2\pi/(5\pm1)a_0$ in the (0, $\pm\pi$) direction (Fig. 3C) and is non-dispersing with energy. Remarkably, this STM detected q-vector matches well with the vector connecting the Fermi arc tip observed in our ARPES data which is $2\pi/(5.2\pm0.7)a_0$ (Fig. 3E) and not the vector connecting the antinodal sections. While the non-dispersive nature of the ordering is clearly visible and rules out the possibility of quasiparticle interference, can this charge order be classified as a bona fide charge density wave (CDW) that can result in a CDW gap? Based on our STM data, we have evidence that this is indeed the case. An important signature of CDW in the local density of states is contrast reversal[20]. Similar to previous STM data, we find no evidence for contrast reversal at low energies (<15 mV bias)[21]. But at higher energies ($\pm$50mV), we observe ubiquitous signals for contrast reversal exemplified in Fig. 4A.

Interestingly, clear contrast reversal in the CDW pattern is also observable as the bias voltage sweeps from 40meV to below 20meV. The coherence length for the CDW pattern is rather short and approximately ~10-15Å, indicative of a glassy CDW, which would make it difficult to observe using scattering probes.

In order to further explore the connection of the CDW with the antinodal ARPES gap and $\Delta_L$ we consider the over-doped 0.1La samples where the vector connecting the arc tip observed by ARPES is expected to result in a spatial ordering with periodicity ~26Å (Fig. 3F). Since the CDW displays short range coherence, this larger periodicity is not likely to be sustainable. More importantly, the kinetic motion of the increasing number of doped holes is enhanced in the overdoped samples. Correspondingly, we find no clear CDW pattern at this doping (Fig. 3, B and D) and the large gap is suppressed in magnitude (Fig. 1, G and H). The concomitant suppression of the large gap and the CDW pattern suggests an intimate connection between the two, even implying a likely causal relationship. The weak pseudogap in the overdoped samples could potentially arise either from CDW fluctuations or a static CDW that is too weak or disordered to be observed by STM.

Finally, CDW ordering is just one amongst many explanations for this pseudogap and while we cannot entirely rule out the possibility that the occurrence of the CDW is coincidental or a surface phenomenon, we can certainly comment on an orthogonal explanation i.e., the fluctuating pair origin of this pseudogap. Based on our data it would be difficult to reconcile our observations of two coexisting gaps below $T_c$ with the assumption that the observed large pseudogap is caused by fluctuating (precursor) pairs. In that case, only a single gap below $T_c$ is expected both in real space and in momentum space. We would like to stress however, that our data on optimal and overdoped samples do not rule out the possibility of an additional smaller pseudogap related to pair fluctuation existing for a narrow temperature range above $T_c$.

An important observation in the STM data is that there is a strong correlation between the small and the large gaps. Figure 4C shows STM spectra on 0.4La samples that were sorted and averaged based on $\Delta_L$. One can clearly see that as the large gap increases, so does the magnitude of the small gap. This is shown to be true statistically as well, as seen by the scatter plot in figure 4B. We have calculated the cross correlation between the two gaps and find that the onsite correlation coefficient is 0.6, which is rather strong, and retains a finite value up to a few lattice constants (Fig. 4B inset). The correlation between the two gaps indicates that both order parameters are influenced by the same underlying physical processes. A key point to note here is that while the second larger gap in the 0.1La samples is rather weak and not always observed, wherever it is visible it shows the same correlation with the superconducting gap as the 0.4La sample (Fig. 4B). Furthermore, the seamless continuation of the scatter plot for the 0.1La samples with the 0.4La samples suggests that the gap variations for both gaps may arise from doping inhomogeneities[22, 23, 24].

In summary, the STM and ARPES data reveal critical new information about the pseudogap phase in Bi2201. Firstly, unlike Bi2212, the pseudogap and Fermi arc above $T_c$ extend into the over-doped regime. Secondly, in Bi2201, a pseudogap above $T_c$ is accompanied by two gaps below $T_c$ that coexist in real space and in momentum space, clearly leading us towards the competing order explanation for this elusive phase.

# Figure 1

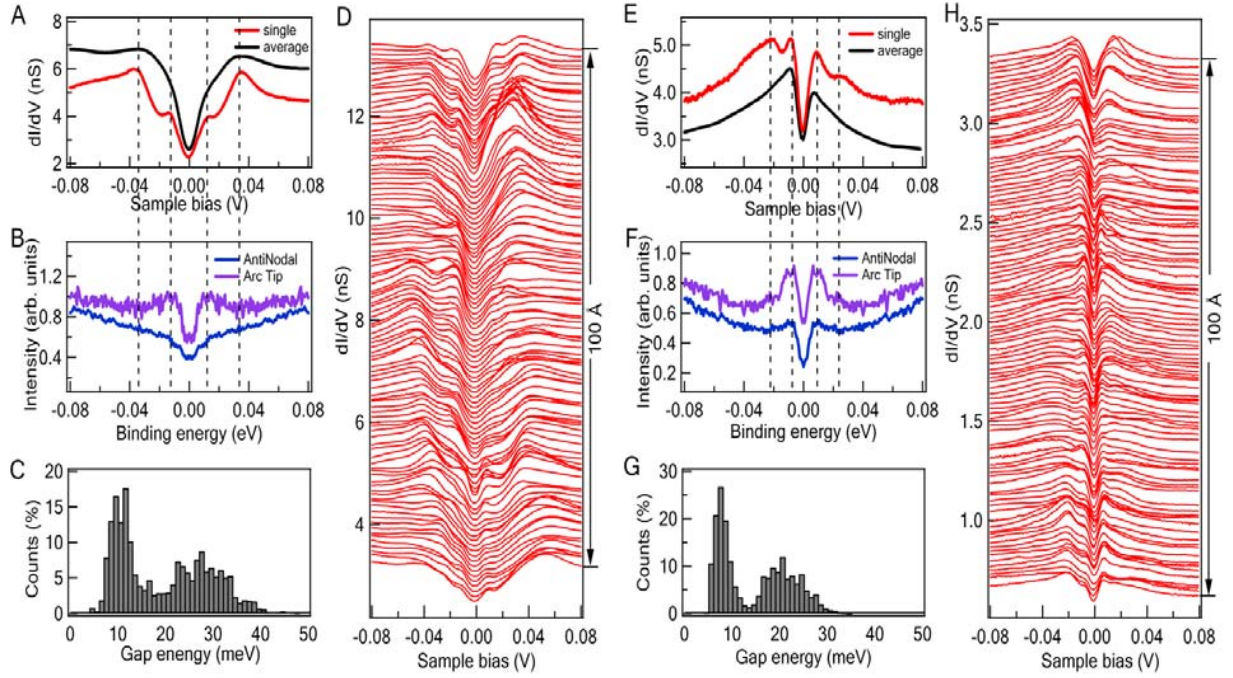

**Fig. 1** Spectroscopic features of nearly optimally doped La-Bi2201with $T_c$~32K (x=0.4) and overdoped La-Bi2201with $T_c$~16K (x=0.1). (**A**) Single STM spectrum (red) representative of the average small gap (set up conditions: sample-bias voltage ($V_s$) = -110mV and a tunnel current ($I_t$) = 500pA) and a spatially averaged STM spectrum (black) from a 240Å map ($V_s$ = -86 mV $I_t$=500pA) for 0.4La sample. (**B**) Symmetrized ARPES energy distribution curves (EDCs) of 0.4La taken at antinodal position and at the arc tip. (**C**) Gap histogram for 0.4La sample from a 237 Å *dI/dV* map showing a bimodal distribution with the small gap ($\Delta_S$) average at 10.50 meV± 2.8 meV and large gap ($\Delta_L$) average at 27.2 meV ± 5.4 meV. (**D**) STM spectra along a line (linecut) over a 100 Å length ($V_s$=-110 mV, $I_t$ =500pA). (**E**) Single STM spectrum for 0.1La sample representative of the average small gap ($V_s$=-100 mV, $I_t$ =400pA) and average spectrum is from a 110 Å map ($V_s$=-100 mV, $I_t$ =300pA). (**F**) Symmetrized ARPES EDCs of 0.1 sample taken at antinodal and arc tip. (**G**) Gap histogram

for 0.1 sample from a 110Å *dI/dV* map ($V_s$=-110 mV, $I_t$ =400pA). Average $\Delta_s$ is 7.4 meV ± 1.6 meV and average $\Delta_L$ is 20.7 meV ± 3.9 meV. (**H**) STM linecut over 100 Å length ($V_s$=-100mv, $I_t$=400pA).

## Figure 2

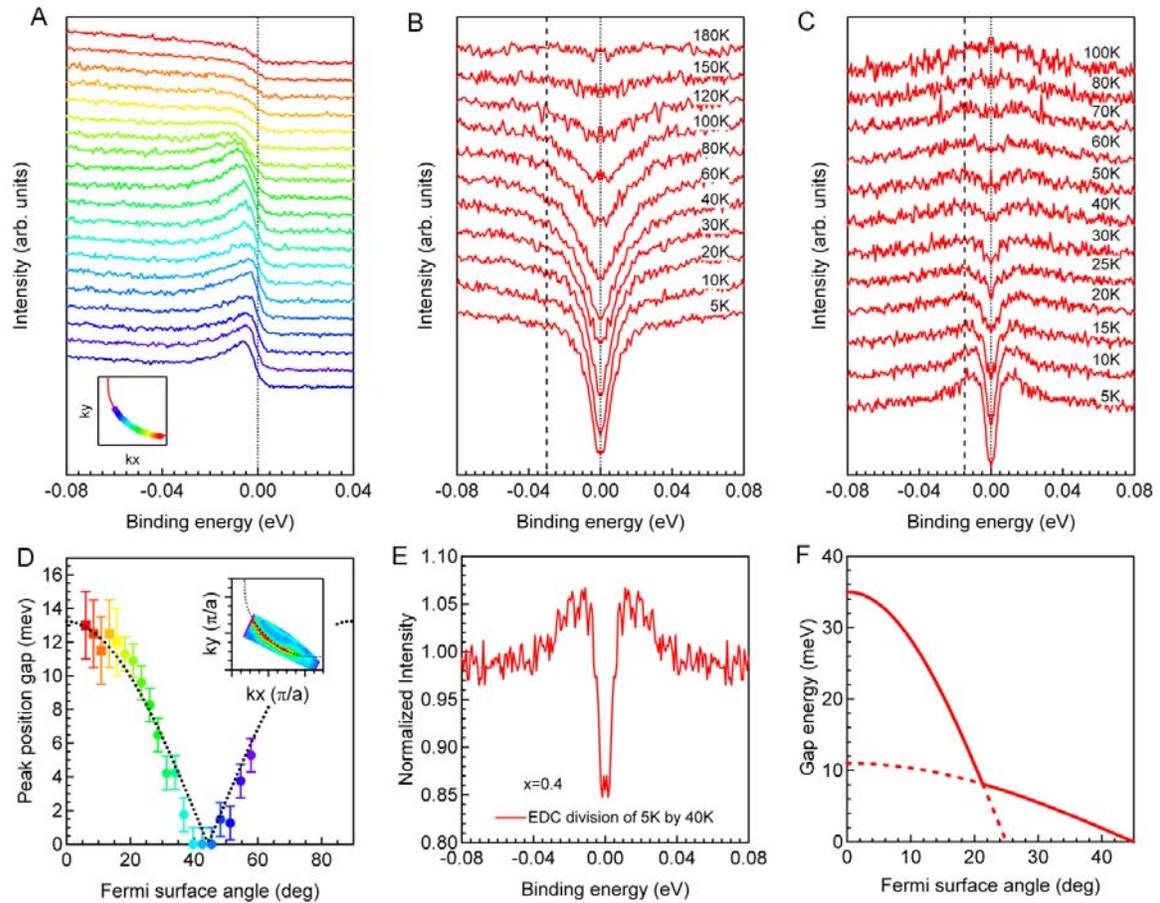

**Fig. 2** ARPES EDCs showing momentum and temperature dependence of gap for 0.4La and 0.1La samples. (**A**) Momentum (k) dependence of EDCs of 0.4La sample used to extract the gap magnitudes shown in **D**. (**B**) Temperature dependence of symmetrized EDCs for 0.4La sample at antinode. Dotted line at 30meV is a guide to the eye to show the large pseudogap observed above $T_c$ (60K or 80K for example). (**C**) Temperature dependence of symmetrized EDCs for 0.1La sample at antinode. (**D**) k-dependent gap value *Vs* angle for 0.4 sample extracted by locating the coherence peak position (dots), a feasible method only up to the coherence arc tip, and by the position of the slope change thereafter (squares). (**E**) Division of antinodal 5K EDC by 40K EDC for 0.4La sample. (**F**) Schematic of gap *Vs* angle showing

the previously proposed two gap picture (solid lines) and our new proposal (solid plus dotted lines) based on the STM and ARPES data.

# Figure 3

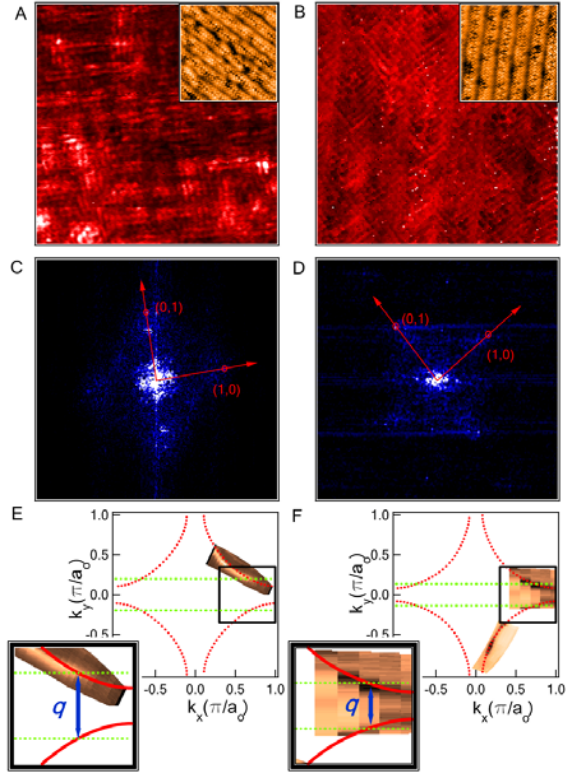

Fig. 3 STM spatial maps, Fourier transform and ARPES Fermi surface. (**A**) 160Å spatial conductance map [g(r,E)] of 0.4La sample at energy (E) = +7 meV ($V_s$ = +120mV, $I_t$=500pA). Inset shows corresponding topographic image taken in the same field of view as the map. (**B**) 160Å g(r,E) map of 0.1 sample at E = -11meV ($V_s$ = -120mV, $I_t$=200pA) inset is the topographic image taken in the same field of view with the map. **C**, Fourier transform of 185 Å map (includes region shown in A) on 0.4La sample showing the q-vectors arising from the CDW pattern shown in (A). The unit for the axis is $2\pi/a_0$ where lattice constant $a_0$ = 3.83Å. (**D**) Fourier transform of map B. The unit for the axis is $2\pi/a_0$. (**E**) ARPES Fermi surface mapping of 0.4La sample. The inset shows the nesting vector at the arc tip $\left(q \approx 2\pi/(5.2\pm0.7)a_0\right)$ which matches the average STM periodicity. (**F**) Fermi surface

mapping of 0.1La sample from ARPES showing corresponding q-vector $\left(q \approx 2\pi/(7\pm0.5)a_0\right)$ which is smaller than the 0.4La sample.

# Figure 4

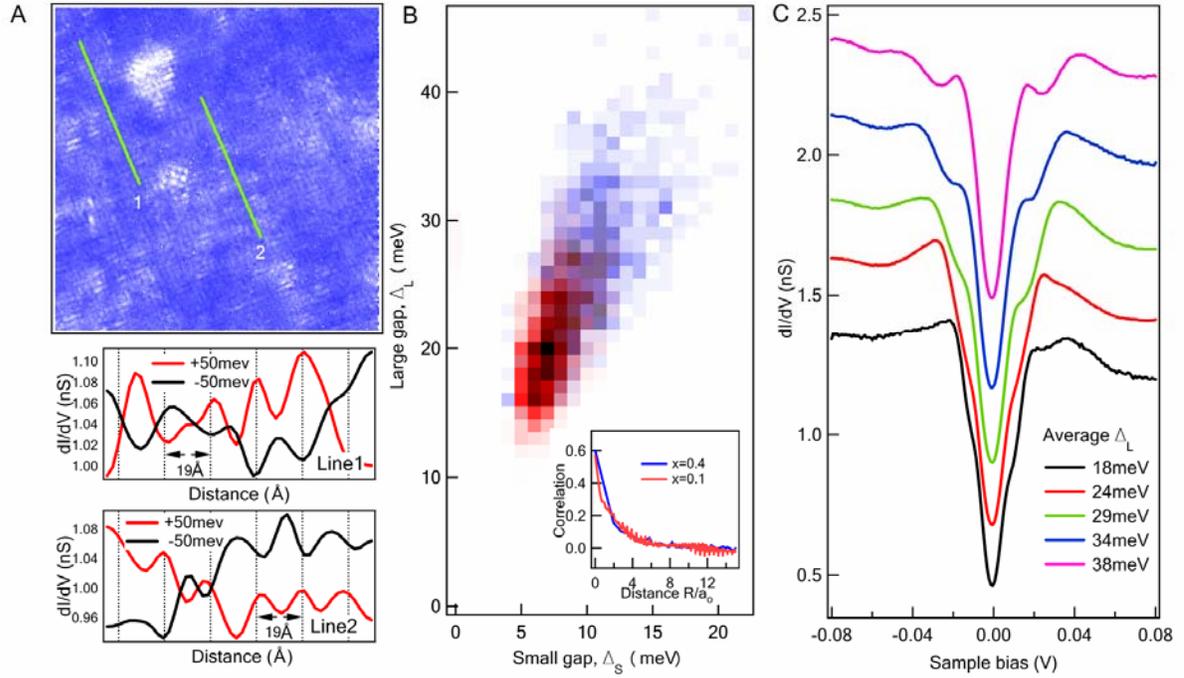

**Fig. 4** STM data illustrating contrast reversal of CDW at opposite bias voltages and the correlation between large and small gaps. **(A)** 237Å g(r,E) map at E = 0 meV, ($V_s$ = -70mV, $I_t$=100pA) showing the position of the two linecuts (Line 1 and Line 2) which are used to illustrate contrast reversal. Shown below the map, are linecuts from E = +50 meV and -50 meV *dI/dV* maps obtained simultaneously with the map shown in A. The linecuts were smoothed 5 times (binomial smoothing) to remove the atomic scale features. The approximate peak to peak distance is around 19Å consistent with the observed CDW periodicity of ~$5a_0$. **(B)** A scatter plot of occurrences of $\Delta_s$ and $\Delta_L$ for the data used to obtain the histograms of Fig. 1(D) (blue squares) and Fig. 1(H) (red squares). The inset shows the spatial dependence of the cross correlation between the $\Delta_s$ and $\Delta_L$ maps, for both 0.4 and 0.1 samples. **(C)** Average of spectra from a 240Å map sorted on the basis of $\Delta_L$ ($V_s$ = -86 mV,

$I_t$=500pA). The corresponding $\Delta_s$ values for each of these averaged spectra are 8 meV, 10 meV, 12 meV, 15 meV and 18 meV respectively.